\documentstyle[pra,aps,psfig]{revtex}
\def\Re{\mathop{\rm Re}} 
\def\Im{\mathop{\rm Im}} 

\begin{document}
\title{Casimir force between partially transmitting mirrors}
\author{Marc Thierry Jaekel $^{(a)}$ and Serge Reynaud $^{(b)}$}
\address{(a) Laboratoire de Physique Th\'{e}orique de l'ENS
\thanks{%
Unit\'e propre du Centre National de la Recherche Scientifique,
associ\'ee \`a l'Ecole Normale Sup\'erieure et \`a l'Universit\'e
Paris-Sud}, 24 rue Lhomond, F75231 Paris Cedex 05 France\\
(b) Laboratoire de Spectroscopie Hertzienne
\thanks{%
Unit\'e de l'Ecole Normale Sup\'erieure et de l'Universit\'e
Pierre et Marie Curie, associ\'ee au Centre National de la Recherche 
Scientifique}, 4 place Jussieu, case 74, F75252 Paris Cedex 05 France}
\date{{\sc Journal de Physique I} {\bf 1} (1991) 1395-1409}
\maketitle

\begin{abstract}
The Casimir force can be understood as resulting from the radiation pressure
exerted by the vacuum fluctuations reflected by boundaries. We extend this
local formulation to the case of partially transmitting boundaries by
introducing reflectivity and transmittivity coefficients obeying conditions
of unitarity, causality and high frequency transparency. We show that the
divergences associated with the infiniteness of the vacuum energy do not
appear in this approach. We give explicit expressions for the Casimir force
which hold for any frequency dependent scattering and any temperature. The
corresponding expressions for the Casimir energy are interpreted in terms of
phase shifts. The known results are recovered at the limit of a perfect
reflection.

PACS: 03.65 - 42.50 - 12.20
\end{abstract}

\section*{Introduction}

As a consequence of Heisenberg inequalities, the electromagnetic field
exhibits quantum fluctuations even in the vacuum state. These ``vacuum
fluctuations'' have been extensively studied in the domain of quantum optics
these last years \cite{Casimir1}. Casimir \cite{Casimir2} observed that the
resulting vacuum energy ($\frac 1  2 \hbar \omega $ per field mode) depends
upon the boundary conditions and particularly upon the positions of
reflecting bodies. As a consequence, the vacuum fluctuations manifest
themselves through macroscopic forces.

These Casimir forces are usually computed by comparing the mode density in
the absence and in the presence of perfectly reflecting boundaries \cite
{Casimir3}. In a local formulation, they can also be understood as resulting
from the radiation pressure exerted by the vacuum fluctuations reflected by
the boundaries \cite{Casimir4,Casimir5}. In the simplest case of a scalar
field in a two-dimensional (2D) spacetime, one gets the following force
between two pointlike mirrors at a distance $q$ 
\begin{equation}
F=\frac{\hbar c\pi }{24q^2 }  \label{eq1}
\end{equation}
For the electromagnetic field in a four-dimensional (4D) spacetime, one gets
the Casimir pressure (force measured per unit area) between two parallel
plane and infinite mirrors \cite{Casimir6} 
\begin{equation}
F=\frac{\hbar c\pi ^2 }{240q^{4}}  \label{eq2}
\end{equation}
From now on, we use natural units where $c=\varepsilon _0 =k_{\rm B}=1$;
however, we keep $\hbar $ as a scale for vacuum fluctuations.

A problem in any calculation of vacuum induced effects is to dispose of the
divergences associated with the infiniteness of the total vacuum energy.
Usually, this is done by arbitrarily cutting off the mode density. The
divergences of the energy in the absence and in the presence of the
boundaries cancel each other, which leads to a finite net result at the
limit of infinite cutoff frequency. In the local formulation, the field
correlation function diverges when the fields are evaluated at the same
point. A regularized stress tensor is obtained by splitting the two points
and ignoring the divergent part \cite{Casimir4,Casimir7}.

Clearly, a more natural regularization should be provided by studying
partially transmitting mirrors. Any real mirrors are certainly transparent
at high frequencies so that the expression of the force should be regular.
Using the physical properties of a dielectric constant, Lifshitz \cite
{Casimir8} has obtained regular expressions for the Casimir-Polder forces
between macroscopic dielectric bodies \cite{Casimir9,Casimir10}. This
approach is not free from problems, such as the effect of dissipation and
the expression of energy density in a dispersive dielectric medium, which
have been discussed in detail \cite{Casimir11,Casimir12}.

In this paper, we use a scattering approach to avoid these problems. We use
the local formulation and introduce frequency dependent reflectivity and
transmittivity coefficients for the two mirrors, which are assumed to obey
conditions of reality, unitarity, causality and high frequency transparency.
The scattering coefficients determine the vacuum stress tensor and therefore
the Casimir force on the mirrors. Explicit expressions, which come out as
finite integrals, are obtained for any frequency dependent scattering.
Eventually, the reflectivity function appears as a physical regulator and
the known results (\ref{eq1}) and (\ref{eq2}) are recovered when the mirrors
are totally reflecting over a large enough frequency interval.

These results do not rely upon a detailed microscopic analysis of the
scattering process and are not limited to dielectric mirrors. Dissipative
processes are disregarded (unitary scattering) and all field fluctuations
originate from the input vacuum. We also give the results for a non zero
temperature input state. The Casimir energy can be deduced by integrating
the force and is different from the integrated field energy when the
reflectivities are frequency dependent. This expression, interpreted in
terms of phase shifts, makes the connection with the mode density approach.

A first part details the simple case where two pointlike mirrors are placed
in the vacuum state, or in a thermal state, of a scalar field in a 2D
spacetime. We come then to the problem of parallel plane and infinite
mirrors which scatter the electromagnetic field in a 4D spacetime. Following
Lifshitz \cite{Casimir8}, the effect of evanescent waves is accounted for.
Expressions of the forces as integrals over imaginary frequencies \cite
{Casimir8} are given in an appendix.

The following convention is used throughout the paper: any function $f(t)$
defined in the time domain and its Fourier transform $f[\omega ]$ are
related through \footnote{{\it The notation used in the original paper for
Fourier transforms has been changed to a more convenient one.}} 
\[
f(t) =\int \frac{{\rm d}\omega }{2\pi }f[\omega ]e^{-i\omega t}
\]

\section*{One mirror in 2D vacuum}

In a 2D spacetime (one time coordinate $t$, one space coordinate $x$), a
free field is the sum of two counterpropagating fields 
\[
\Phi =\varphi \left( t-x\right) +\psi \left( t+x\right) 
\]
It can also be described by a twofold column matrix in the frequency domain 
\[
\Phi [\omega ]=\left( 
\begin{array}{c}
\varphi [\omega ] \\ 
\psi [\omega ]
\end{array}
\right) 
\]

We consider now that the field is scattered by a mirror located at point $q$ 
\[
\Phi =\theta (q-x)\left( \varphi _{\rm in}(t-x)+\psi _{\rm out}
(t+x)\right) +\theta (x-q)\left( \psi _{\rm in}(t+x)+\varphi _{\rm out}
(t-x)\right) 
\]
where $\varphi _{\rm in}$ and $\psi _{\rm in}$ are the input fields, 
$\varphi _{\rm out}$ and $\psi _{\rm out}$ the output fields 
(see Figure 1). 
\begin{figure}[h]
\centerline{\psfig{figure=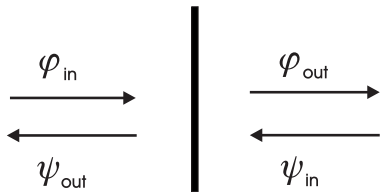,width=5cm}}
\caption{One mirror scatters the two counterpropagating fields.}
\label{Fig1}
\end{figure}
With the hypothesis of a perfect reflection, the field vanishes on the
mirror 
\[
\varphi _{\rm out}(t)=-\psi _{\rm in}(t+2q)\qquad \psi _{\rm out}
(t)=-\varphi _{\rm in}(t-2q)
\]
Equivalently 
\[
\varphi _{\rm out}[\omega ]=-\psi _{\rm in}[\omega ]e^{-2i\omega q}
\qquad \psi _{\rm out}[\omega ]=-\varphi _{\rm in}[\omega ]e^{2i\omega q}
\]
In the general case of a partially transmitting mirror, the scattering of
the field is described by a frequency dependent $S$ matrix 
\begin{equation}
\Phi _{\rm out}[\omega ]=S[\omega ]\Phi _{\rm in}[\omega ]\qquad
S[\omega ]=\left( 
\begin{array}{cc}
s[\omega ] & r[\omega ]e^{-2i\omega q} \\ 
r[\omega ]e^{2i\omega q} & s[\omega ]
\end{array}
\right)   \label{eq3}
\end{equation}

We assume that the $S$ matrix obeys the following conditions. The values 
of $S$ are real in the temporal domain 
\begin{equation}
s[-\omega ]=s[\omega ]^{*}\qquad r[-\omega ]=r[\omega ]^{*}  \label{eq4}
\end{equation}
The scattering is causal \cite{Casimir13} 
\begin{eqnarray}
&&s(t)=r(t)=0 \ {\rm for}\ t<0  \label{eq5} \\
&&s[\omega ]\ {\rm and}\ r[\omega ]\ {\rm are\ analytic\ and\ regular\ for}
\ \Im \omega >0  \nonumber
\end{eqnarray}
The scattering matrix is unitary (dissipation inside the mirror is
neglected) 
\begin{eqnarray}
&&S[\omega ]S[\omega ]^\dagger =1  \nonumber \\
\left| s[\omega ]\right| ^2 +\left| r[\omega ]\right| ^2  &=&1\qquad
s[\omega ]r[\omega ]^{*}+r[\omega ]s[\omega ]^{*}=0  \label{eq6}
\end{eqnarray}
The mirror is transparent at high frequencies and the reflectivity decreases
sufficiently rapidly 
\begin{equation}
s[\omega ]\rightarrow 1\qquad \omega \left| r[\omega ]\right| \rightarrow
0\qquad {\rm for}\ \omega \rightarrow \infty   \label{eq7}
\end{equation}

Because of the transparency condition (\ref{eq7}), the reflectivity will
provide a natural regulator to the ultraviolet divergence of the vacuum
energy. The perfectly reflecting mirror corresponds to $r=-1$ and $s=0$ at
all frequencies and does not obey this condition. So, it will be preferable
to consider the perfect mirror as the limit of a model obeying the
conditions (\ref{eq4}-\ref{eq7}). This amounts to consider the effect of non
zero reflection delays and to let them go to zero at the end of the
calculations.

Quantum field theory provides the field correlation functions corresponding
to the vacuum state. We will use the symmetric correlation functions which
are the mean values of the anticommutators 
\begin{eqnarray}
\left\langle \left\{ \varphi ^\prime (t+\tau ),\varphi ^\prime
(t)\right\} \right\rangle  &=&\left\langle \left\{ \psi ^\prime (t+\tau
),\psi ^\prime (t)\right\} \right\rangle =\frac{c(t)}2   \nonumber \\
\left\langle \left\{ \varphi ^\prime (t+\tau ),\psi ^\prime (t)\right\}
\right\rangle  &=&0  \nonumber \\
\varphi ^\prime (t) &=&\partial _{t}\varphi (t)  \label{eq8} \\
i\omega i\omega ^\prime \left\langle \left\{ \varphi [\omega ],\varphi
(\omega ^\prime )\right\} \right\rangle  &=&i\omega i\omega ^\prime
\left\langle \left\{ \psi [\omega ],\psi (\omega ^\prime )\right\}
\right\rangle =2\pi \delta (\omega +\omega ^\prime )\frac{c[\omega ]}2  
\nonumber \\
i\omega i\omega ^\prime \left\langle \left\{ \varphi [\omega ],\psi
(\omega ^\prime )\right\} \right\rangle  &=&0  \nonumber
\end{eqnarray}
with 
\begin{eqnarray}
c(\tau ) &=&\lim_{\epsilon \rightarrow 0^{+}}\frac{\hbar }{2\pi }\left[ 
\frac 1  {\left( \epsilon +i\tau \right) ^2 }+\frac 1  {\left( \epsilon
-i\tau \right) ^2 }\right]   \nonumber \\
c[\omega ] &=&\hbar \left| \omega \right|   \label{eq9}
\end{eqnarray}

As usual in the local formulation, one evaluates the force as the difference
between the radiation pressures exerted upon the left and right sides of the
mirror. In a 2D spacetime, the component $T_{xx}$ of the stress tensor is
equal to the energy density and one gets 
\begin{eqnarray}
&&F=e_{\rm L}-e_{\rm R}  \nonumber \\
e_{\rm L} &=&\left\langle \varphi _{\rm in}^{\prime \ 2}(t-q)+\psi _
{\rm out}^{\prime \ 2}(t+q)\right\rangle \qquad e_{\rm R}=\left\langle
\psi _{\rm in}^{\prime \ 2}(t+q)+\varphi _{\rm out}^{\prime \
2}(t-q)\right\rangle   \label{eq10}
\end{eqnarray}
Substituting the output fields in terms of the input fields and using the
unitarity condition, one obtains 
\[
\left\langle \varphi _{\rm out}^{\prime \ 2}(t-q)\right\rangle
=\left\langle \psi _{\rm out}^{\prime \ 2}(t+q)\right\rangle =\left\langle
\varphi _{\rm in}^{\prime \ 2}(t-q)\right\rangle =\left\langle \psi _
{\rm in}^{\prime \ 2}(t+q)\right\rangle =\frac{c(0)}{4}
\]
Hence 
\begin{equation}
e_{\rm L}=e_{\rm R}=\int_0 ^{\infty }\frac{{\rm d}\omega }{2\pi }
c[\omega ]  \label{eq11}
\end{equation}
The radiation pressures exerted upon the two sides of the mirror are
infinite. However, they cancel each other exactly and the net force is zero.
This is connected to the property that the impulsion density is zero in the
vacuum state of a free field while the energy density is infinite. Note that
the cancellation between infinite quantities is obtained inside a single
physical situation and not by comparing two different situations.

\section*{Two mirrors in 2D vacuum}

We study now the situation where two pointlike mirrors are placed in the 2D
vacuum at positions $q_1 $ and $q_2 $ (see Fig.2). 
\begin{figure}[h]
\centerline{\psfig{figure=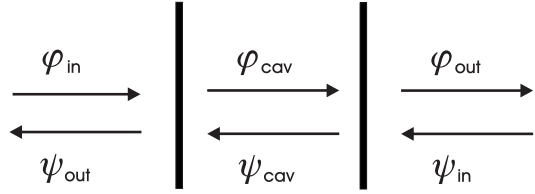,width=7cm}}
\caption{Two mirrors scatter the two
counterpropagating fields.}
\label{Fig2}
\end{figure}
As is usual in the theory of a Fabry Perot cavity, the global scattering may
be deduced from the elementary scattering matrices given by (\ref{eq3}) and
associated with the two mirrors 
\begin{eqnarray*}
\left( 
\begin{array}{c}
\varphi _{\rm cav}[\omega ] \\ 
\psi _{\rm out}[\omega ]
\end{array}
\right)  &=&\left( 
\begin{array}{cc}
s_1 [\omega ] & r_1 [\omega ]e^{-2i\omega q_1 } \\ 
r_1 [\omega ]e^{2i\omega q_1 } & s_1 [\omega ]
\end{array}
\right) \left( 
\begin{array}{c}
\varphi _{\rm in}[\omega ] \\ 
\psi _{\rm cav}[\omega ]
\end{array}
\right)  \\
\left( 
\begin{array}{c}
\varphi _{\rm out}[\omega ] \\ 
\psi _{\rm cav}[\omega ]
\end{array}
\right)  &=&\left( 
\begin{array}{cc}
s_2 [\omega ] & r_2 [\omega ]e^{-2i\omega q_2 } \\ 
r_2 [\omega ]e^{2i\omega q_2 } & s_2 [\omega ]
\end{array}
\right) \left( 
\begin{array}{c}
\varphi _{\rm cav}[\omega ] \\ 
\psi _{\rm in}[\omega ]
\end{array}
\right) 
\end{eqnarray*}
One can solve these equations to express the outcoming and the intracavity
fields in terms of the input ones 
\begin{eqnarray*}
\Phi _{\rm out}[\omega ] &=&S[\omega ]\Phi _{\rm in}[\omega ] \\
\Phi _{\rm cav}[\omega ] &=&R[\omega ]\Phi _{\rm in}[\omega ]
\end{eqnarray*}
The global scattering matrix $S[\omega ]$ and the resonance matrix 
$R[\omega]$ are given by 
\begin{eqnarray}
S_{11}[\omega ] &=&S_{22}[\omega ]=\frac{s_1 [\omega ]s_2 [\omega ]}
{d[\omega ]}  \nonumber \\
S_{12}[\omega ] &=&r_2 [\omega ]e^{-i\omega q}+\frac{s_2 [\omega
]^2 r_1 [\omega ]e^{i\omega q}}{d[\omega ]}  \nonumber \\
S_{21}[\omega ] &=&r_1 [\omega ]e^{-i\omega q}+\frac{s_1 [\omega
]^2 r_2 [\omega ]e^{i\omega q}}{d[\omega ]}  \label{eq12}
\end{eqnarray}
\begin{eqnarray}
R_{11}[\omega ] &=&\frac{s_1 [\omega ]}{d[\omega ]}\qquad R_{12}[\omega ]=
\frac{s_2 [\omega ]r_1 [\omega ]e^{-2i\omega q_1 }}{d[\omega ]}  \nonumber
\\
R_{21}[\omega ] &=&\frac{s_1 [\omega ]r_2 [\omega ]e^{2i\omega q_2 }}
{d[\omega ]}\qquad R_{22}[\omega ]=\frac{s_2 [\omega ]}{d[\omega ]}
\label{eq13}
\end{eqnarray}
where 
\[
d[\omega ]=1-r[\omega ]e^{2i\omega q}\qquad r[\omega ]=r_1 [\omega
]r_2 [\omega ]\qquad q=q_2 -q_1 
\]
The matrices $S$ and $R$ are real in the time domain, they are causal
functions and the matrix $S$ is unitary.

Substituting the output fields in terms of the input fields and using the
unitarity of the matrix $S[\omega ]$, one deduces that the energy densities 
$e_{\rm L}$ and $e_{\rm R}$, again defined by (\ref{eq10}), are still
given by (\ref{eq11}). But the intracavity energy density $e_{\rm cav}$ is
different 
\begin{eqnarray*}
e_{\rm cav} &=&\left\langle \varphi _{\rm cav}^{\prime \ 2}+\psi _{\rm cav}
^{\prime \ 2}\right\rangle =\int_0 ^{\infty }\frac{{\rm d}\omega }
{2\pi }c[\omega ]g[\omega ] \\
g[\omega ] &=&\frac{\left| R_{11}\right| ^2 +\left| R_{12}\right|
^2 +\left| R_{21}\right| ^2 +\left| R_{22}\right| ^2 }2 =\frac{1-\left|
r[\omega ]\right| ^2 }{\left| 1-r[\omega ]e^{2i\omega q}\right| ^2 }
\end{eqnarray*}
The function $g[\omega ]$ (calculated from equation (\ref{eq13})) describes
an enhancement (resp. suppression) of vacuum fluctuations seen inside the
cavity at frequencies inside (resp. outside) the Airy peaks \cite
{Casimir14,Casimir15}. One deduces the Casimir force as the difference
between the outer and inner radiation pressures 
\begin{equation}
F_1 =e_{\rm L}-e_{\rm cav}=F\qquad F_2 =e_{\rm cav}-e_{\rm R}=-F
\label{eq14}
\end{equation}
\begin{equation}
F=\int_0 ^{\infty }\frac{{\rm d}\omega }{2\pi }c[\omega ]\left( 1-g[\omega
]\right)   \label{eq15}
\end{equation}

In the standard derivation of the Casimir force, one considers mirrors which
are perfectly reflecting at all frequencies. In this case, the Airy peaks
reduce to Dirac distributions 
\[
1-g[\omega ]=1-\frac{\pi }{q}\sum_{n}\delta (\omega -n\frac{\pi }{q})
\]
The integral (\ref{eq15}) is then regularized by introducing a cutoff in the
mode density and computed by applying the Euler-Mac Laurin summation rule 
\cite{Casimir6} 
\[
F=\left\{ \left( \frac{B_2 }{2!}\frac{\pi ^2 }{q^2 }\partial _\omega +
\frac{B_{4}}{4!}\frac{\pi ^{4}}{q^{4}}\partial _\omega ^{3}+\ldots \right) 
\frac{\hbar \omega }{2\pi }\right\} _{\omega =0}
\]
where the $B_{2k}$ are the Bernouilli numbers $B_2 =\frac 1 {6};$ 
$B_{4}=- \frac 1  {30};$ $\ldots $. This leads to the expression (\ref{eq1}) of the
Casimir force.

However, the introduction of a cutoff in the mode density is not necessary
since the integral (\ref{eq15}) is finite as soon as the reflectivity
functions obey the conditions (\ref{eq4}-\ref{eq7}). One can write 
\begin{equation}
F=\int_0 ^{\infty }\frac{{\rm d}\omega }{2\pi }c[\omega ]\frac{-r[\omega
]e^{2i\omega q}}{1-r[\omega ]e^{2i\omega q}}+c.c.  \label{eq16}
\end{equation}
As $c[\omega ]$ is an even function of $\omega $, one deduces from the
reality condition (\ref{eq4}) 
\begin{equation}
F=\int_{-\infty }^{\infty }\frac{{\rm d}\omega }{2\pi }c[\omega ]\frac{
-r[\omega ]}{e^{-2i\omega q}-r[\omega ]}  \label{eq17}
\end{equation}
One obtains the force as a sum 
\begin{equation}
F=\sum_{\ell >0}F^{(\ell )}\qquad F^{(\ell )}=-\int_{-\infty }^{\infty }
\frac{{\rm d}\omega }{2\pi }e^{2\ell i\omega q}r^{\ell }[\omega ]c[\omega ]
\label{eq18}
\end{equation}
where each term $F^{(\ell )}$ corresponds to a number $\ell $ of roundtrips
inside the cavity. This term can be transformed into a convolution product
in the time domain, more precisely into an integral over $\ell $ reflection
delays $t_1 $, $\ldots $ $t_{\ell }$ 
\begin{equation}
F^{(\ell )}=-\int_0 ^{\infty }{\rm d}t_1 r(t_1 )\ldots \int_0 ^{\infty }
{\rm d}t_{\ell }r(t_{\ell })c\left( 2\ell q+t_1 +\ldots +t_{\ell }\right) 
\label{eq19}
\end{equation}
The reflection delays are positive and the two point correlation function $c$
is evaluated at distances between the points greater than $2q$. The
expression of the force comes out free from the divergences usually
associated with the infiniteness of the vacuum energy and the regulator 
$\epsilon $ may be omitted in the expression (\ref{eq9}) of $c$.

In the limiting case where the reflection delays ($t_1 $, $\ldots $ 
$t_{\ell }$) are much shorter than the distance $q$ between the two mirrors
(large distance approximation), one can neglect them in the function 
$c(2\ell q+t_1 +\ldots +t_{\ell })$. It follows that the force depends only
upon the reflectivities evaluated at zero frequency 
\[
F=-\sum_{\ell >0}r_0 ^{\ell }c(2\ell q)\qquad r_0 =r[0]=r_1 [0]r_2 [0]
\]
that is 
\begin{equation}
F=\frac{\hbar }{4\pi q^2 }\zeta _{r_0 }(2)  \label{eq20}
\end{equation}
\begin{equation}
\zeta _{x}(p)=\sum_{\ell =1}^{\infty }\frac{x^{\ell }}{\ell ^{p}}
\label{eq21}
\end{equation}
If the two mirrors are perfectly reflecting at zero frequency, one finds the
expression (\ref{eq1}) for the Casimir force $\left( \zeta _1 (2)=\frac{\pi
^2 }{6}\right) $.

The equation (\ref{eq18}) allows one to evaluate the correction to the limit
of ideal mirrors. It can be written as a formal series 
\begin{eqnarray}
F^{(\ell )} &=&\frac{\hbar }{2\pi }\left\{ -i\left( 2\ell q-i\partial
_\omega \right) ^{-1}\left( \omega r[\omega ]^{\ell }\right) \right\}
_{\omega =0}+c.c.  \nonumber \\
&=&\frac{\hbar }{2\pi }\left\{ \left( 2\ell q-i\partial _\omega \right)
^{-2}r[\omega ]^{\ell }\right\} _{\omega =0}+c.c.  \label{eq22}
\end{eqnarray}
When $r[\omega ]$ is a very flat function, one recovers the large distance
approximation where the force depends upon the value $r_0 $ evaluated at
zero frequency. The terms of the formal series which depend upon the
derivatives of $r[\omega ]$ may be considered as corrections to this
approximation.

These discussions show that the reflectivity function $r[\omega ]$ plays the
same role as a regulator for the mode density. However, it is a regulator
obeying the conditions (\ref{eq4}-\ref{eq7}) which cannot be real for real
frequencies. We show in the appendix that it is possible to write the force
as an integral over imaginary frequencies and that $r[i\omega ]$ appears as
a real regulator \cite{Casimir8}.

\section*{Casimir energy}

The local formulation led us to the expression of the force. One can deduce
a Casimir energy ${\cal U}$ by integrating this force (see equation (\ref
{eq14})) 
\begin{equation}
{\rm d}{\cal U}=-F_1 {\rm d}q_1 -F_2 {\rm d}q_2 =F{\rm d}q  \label{eq23}
\end{equation}
We want now to compute this Casimir energy and to compare it with the
integrated field energy \cite{Casimir4}.

One deduces from equation (\ref{eq16}) 
\begin{eqnarray}
F &=&\partial _{q}{\cal U\qquad U}=\int_0 ^{\infty }\frac{{\rm d}\omega }
{2\pi }\frac{-\hbar \Delta [\omega ]}2   \label{eq24} \\
&&\Delta [\omega ]=i{\rm Log}\frac{1-r[\omega ]e^{2i\omega q}}{1-r[\omega
]^{*}e^{-2i\omega q}}  \label{eq25}
\end{eqnarray}
One deduces respectively from the reality condition (\ref{eq4}) and from the
transparency condition (\ref{eq7}) 
\[
\Delta [0]=0\qquad \omega \Delta [\omega ]\rightarrow 0\qquad {\rm for\ }
\omega \rightarrow \infty 
\]
so that the energy can be written 
\begin{equation}
{\cal U}=\int_0 ^{\infty }\frac{{\rm d}\omega }{2\pi }\frac{\hbar \omega }
2 \partial _\omega \Delta [\omega ]  \label{eq26}
\end{equation}
This is the standard form for the phase shift representation of the Casimir
energy \cite{Casimir3}. Actually, one checks from the expression (\ref{eq12}) 
of the $S$ matrix that 
\[
\det S[\omega ]=\det S_1 [\omega ]\det S_2 [\omega ]e^{i\Delta [\omega ]}
\]
where $S_1 [\omega ]$ and $S_2 [\omega ]$ are the elementary $S$ matrices
associated with the two mirrors; $\Delta [\omega ]$ is the $q-$dependent
part of the sum of the two phase shifts at frequency $\omega $.

Now, one gets from equation (\ref{eq25}) 
\begin{eqnarray}
\partial _\omega \Delta [\omega ] &=&\frac{\left( 2qr[\omega ]-i\partial
_\omega r[\omega ]\right) e^{2i\omega q}}{1-r[\omega ]e^{2i\omega q}}+c.c.
\nonumber \\
&=&-\left( 1-g[\omega ]\right) \left( 2q+\partial _\omega \delta \right)
-g[\omega ]\sin \left( 2\omega q+\delta \right) \partial _\omega {\rm Log}
\left( 1-\rho ^2 \right)   \label{eq27}
\end{eqnarray}
where $\delta $ and $\rho $ are the frequency dependent phase and modulus of
the reflectivity function $r$. Usually, the Casimir effect is computed for
frequency independent reflectivities. In this case, one deduces that the
Casimir energy ${\cal U}$ is identical to the integrated field energy \cite
{Casimir4} 
\[
{\cal U}=q\left( e_{\rm cav}-e_{\rm L}\right) =-qF
\]
This result is compatible with the definition (\ref{eq23}) of ${\cal U}$
since the force scales as $q^{-2}$ in this case \cite{Casimir7}. But the
reflectivities have to be frequency dependent in order to fulfill the
transparency condition (\ref{eq7}). The resulting terms in the Casimir
energy can be considered as corrections associated with the reflection
delays upon the mirrors.

\section*{Casimir force for non zero temperatures}

Now, we derive the expression of the Casimir force for non zero temperatures 
\cite{Casimir3,Casimir4}. We suppose that the reflectivities are independent
of the temperature. Therefore, we have only to change the correlation
function of the input fields.

The Casimir force is given by the integrals (\ref{eq15}) or (\ref{eq17})
with a modified function $c$ 
\begin{equation}
c_{T}[\omega ]=c_0 [\omega ]n_{T}[\omega ]\qquad n_{T}[\omega ]=\frac 1  
{\tanh \left( \frac{\hbar \left| \omega \right| }{2T}\right) }  \label{eq28}
\end{equation}
where $c_0 [\omega ]$ corresponds to the zero temperature and $n_{T}[\omega
]$ is the mean number of thermal photons per mode. The equations (\ref{eq19}) 
are still valid with a function $c_{T}[\omega ]$ which is the Fourier
transform of 
\[
c_{T}(\tau )=-\frac{\hbar }{\pi }\frac{\alpha ^2 }{\sinh ^2 \left( \alpha
\tau \right) }\qquad \alpha =\frac{\pi T}{\hbar }
\]
It is known that the correlation function $c_{T}$ is the same as for an
accelerated vacuum \cite{Casimir16,Casimir17,Casimir18}, which also suggests
that there are relations between the Casimir effect for non zero
temperatures and the physics of black holes \cite{Casimir19}.

At the large distance approximation, one gets 
\[
F=-\sum_{\ell >0}r_0 ^{\ell }c_{T}(2\ell q)
\]
At the low temperature limit, one has the same results as previously. At the
high temperature limit, one obtains an exponentially small force since 
\[
c_{T}(\tau )\approx -\frac{4\pi }{\hbar }T^2 \exp \left( -\frac{2\pi T\tau 
}{\hbar }\right) \ {\rm for\ }T\tau \gg \hbar 
\]
The thermal contribution exactly cancels the vacuum contribution for
temperatures $T$ higher than $\frac{\hbar }{q}$.

In the general case (no approximation on the distance), we obtain the
Casimir free energy ${\cal F}$ by integrating the force \cite{Casimir4} 
\begin{equation}
F=\partial _{q}{\cal F}(q,T)\qquad {\cal F}=\int_0 ^{\infty }\frac{{\rm d}
\omega }{2\pi }\frac{-\hbar \Delta [\omega ]}2 n_{T}[\omega ]  \label{eq29}
\end{equation}
where $\Delta [\omega ]$ is still given by (\ref{eq25}). Noting that (${\cal 
S}$ is the entropy) 
\begin{eqnarray*}
&&{\cal U}={\cal F}+T{\cal S}={\cal F}-T\partial _{T}{\cal F} \\
&&\omega \partial _\omega n_{T}[\omega ]=-T\partial _{T}n_{T}[\omega ]
\end{eqnarray*}
one also obtains 
\[
{\cal U}=\int_0 ^{\infty }\frac{{\rm d}\omega }{2\pi }\frac{-\hbar \Delta
[\omega ]}2 \partial _\omega \left( \omega n_{T}[\omega ]\right) 
\]
An integration by parts then leads to the modified phase shift
representation 
\begin{equation}
{\cal U}=\int_0 ^{\infty }\frac{{\rm d}\omega }{2\pi }\frac{\hbar \omega }
2 n_{T}[\omega ]\partial _\omega \Delta [\omega ]  \label{eq30}
\end{equation}
Using equation (\ref{eq27}), one reaches the same conclusions as previously
about the comparison between the Casimir energy and the integrated field
energy.

\section*{Two mirrors in the electromagnetic vacuum}

We consider now that two parallel, plane and infinite mirrors are placed in
the vacuum state of the electromagnetic field (plate separation $q$ along
the $x$ direction). The scattering on each mirror is still described by a $S$
matrix which obeys conditions of reality, unitarity, causality and high
frequency transparency. Now, the reflection coefficients associated with the
two mirrors depend not only upon the frequency, but also upon the
propagation direction and the polarization. We will denote 
\[
\gamma =\cos \varphi \qquad \kappa =\omega \cos \varphi =\omega \gamma
\qquad \kappa ^\prime =\omega \sin \varphi 
\]
where $\varphi $ is the incidence angle, $\kappa $ the wavevector along the $%
x$ direction and $\kappa ^\prime $ the length of the transverse wavevector.

The calculations of the Casimir force (measured per unit area) are mostly
the same as in the 2D case, but one obtains it as a sum over the two
polarizations $p_1 $ and $p_2 $ 
\[
F=F_{p_1 }+F_{p_2 }
\]
Each contribution $F_{p}$ can be written as an integral over the modes 
\begin{equation}
F_{p}=\int \int \frac{{\rm d}\omega }{2\pi }\omega ^2 \frac{{\rm d}\gamma }
{2\pi }\theta (\omega )c[\omega ]\left( 1-g_{p}[\omega ,\kappa ]\right)
\gamma ^2  \label{eq31}
\end{equation}
The function $c$ is the same as in the 2D case. The function $g_{p}$ is
computed as previously, but its resonances are now determined by the
wavevector measured along $x$ 
\begin{eqnarray*}
g_{p}[\omega ,\kappa ] &=&\frac{1-\left| r_{p}[\omega ,\kappa ]\right| ^2 }
{\left| 1-r_{p}[\omega ,\kappa ]e^{2i\kappa q}\right| ^2 } \\
r_{p}[\omega ,\kappa ] &=&r_{1p}[\omega ,\kappa ]r_{2p}[\omega ,\kappa ]
\end{eqnarray*}
where $r_{1p}$ and $r_{2p}$ are the reflectivities associated with the two
mirrors. In comparison with the 2D computation, there is in equation (\ref{eq31}) 
an extra geometrical factor $\gamma ^2 $ which is the ratio between
the component $T_{xx}$ of the stress tensor (which gives the radiation
pressure upon the mirror) and the energy density \cite{Casimir20}.

If the mirrors are supposed perfectly reflecting at all frequencies, Dirac
distributions appear in the formula (\ref{eq31}) which can be computed by a
regularization and the application of the Euler-Mac Laurin summation rule 
\cite{Casimir6} 
\begin{equation}
F_{p}=\left\{ \left( \frac{B_2 }{2!}\frac{\pi ^2 }{q^2 }\partial _{\kappa
}+\frac{B_{4}}{4!}\frac{\pi ^{4}}{q^{4}}\partial _{\kappa }^{3}+\ldots
\right) \frac{\hbar \kappa ^2 \left( K-\kappa \right) }{4\pi ^2 }\right\}
_{\kappa =0}  \label{eq32}
\end{equation}
The constant $K$ goes to infinity with the cutoff frequency. But the even
derivatives do not appear in the Euler-Mac Laurin sum and this constant is
eliminated, which leads to expression (\ref{eq2}) of the force.

In the general case of a frequency dependent reflectivity, one writes 
\begin{equation}
F_{p}=\int \int \frac{{\rm d}\omega }{2\pi }\frac{{\rm d}\gamma }{2\pi }
\theta \left( \omega \right) c[\omega ]\omega ^2 \gamma ^2 \frac{
-r_{p}[\omega ,\omega \gamma ]}{e^{-2i\omega \gamma q}-r_{p}[\omega ,\omega
\gamma ]}+c.c.  \label{eq33}
\end{equation}
Now, one must specify more precisely the integration path for $\gamma $.
Without further investigation, it could be admitted that it coincides with
the segment $\left[ 0,1\right[ $. Using the same methods as for the 2D
computation, one shows however that this integration path leads to a Casimir
force which diverges at the limit of a perfectly reflecting mirror: there
remains a contribution to the force analogous to the constant $K$ appearing
in (\ref{eq32}).

Actually, the integration path for $\gamma $ must also include the upper
part of the imaginary axis $\left] +i\infty ,0\right] $. Imaginary values of 
$\gamma $ not too far from $0$ correspond to values of $\kappa ^\prime $
greater than, but not too much greater than $\omega $. They are associated
with vacuum fluctuations which enter the system through the side boundaries
of the mirror and which contribute to the field outside the mirrors as
evanescent waves. We will suppose that $r_{p}[\omega ,\kappa ]$ may be
analytically continued \cite{Casimir21} for values of $\kappa $ in the
quadrant $\left[ \Re \kappa >0,\Im \kappa >0\right] $ and also that
the contribution of the evanescent waves to the force is given by the
analytical continuation of the ordinary expression \cite{Casimir22}.
Imaginary values of $\gamma $ farther from $0$ correspond to larger values
of $\kappa ^\prime $. They are associated to ``closed'' input ports and
have actually a zero contribution to the force. Assuming that $\left|
r_{p}[\omega ,\kappa ]\right| $ decreases sufficiently fast for large values
of $\kappa $ and that \cite{Casimir23} 
\[
\left| r_{p}[\omega ,\kappa ]\right| <1\ {\rm for}\ \omega \ {\rm real}\ 
{\rm and}\ \left[ \Re \kappa >0,\Im \kappa >0\right] 
\]
one moves the integration path for $\gamma $ to $\left] 1,\infty \right[ $
so that 
\begin{equation}
F_{p}=\int_0 ^{\infty }\frac{{\rm d}\omega }{2\pi }\int_1 ^{\infty }\frac{
{\rm d}\gamma }{2\pi }c[\omega ]\omega ^2 \gamma ^2 \frac{r_{p}[\omega
,\omega \gamma ]}{e^{-2i\omega \gamma q}-r_{p}[\omega ,\omega \gamma ]}+c.c.
\label{eq34}
\end{equation}

This result is regular at the limit of perfect reflection. As for the 2D
computation, one writes the force (\ref{eq34}) as a sum over a number $\ell $
of roundtrips 
\begin{eqnarray}
F_{p} &=&\sum_{\ell >0}F_{p}^{(\ell )}\qquad F_{p}^{(\ell )}=\int_{-\infty
}^{\infty }\frac{{\rm d}\kappa }{2\pi }e^{2\ell i\kappa q}\hbar \kappa
^2 R_{p}^{(\ell )}[\kappa ]  \nonumber \\
&&R_{p}^{(\ell )}[\kappa ]=\int_0 ^{\kappa }\frac{{\rm d}\omega }{2\pi }
r_{p}[\omega ,\kappa ]^{\ell }  \label{eq35}
\end{eqnarray}
(we have used a reality condition $r_{p}[\omega ,\kappa ]^{*}=r_{p}[-\omega
,-\kappa ]$). This expression can be written as a formal series 
\[
F_{p}^{(\ell )}=\frac{\hbar }{2\pi }\left\{ i\left( 2\ell q-i\partial
_{\kappa }\right) ^{-1}\left( \kappa ^2 R_{p}^{(\ell )}[\kappa ]\right)
\right\} _{\kappa =0}+c.c.
\]
At the limit of a large separation, we retain the terms which depend upon
the reflectivities evaluated around zero frequency but not those which
depend upon the derivatives of the reflectivities. As 
\[
R_{p}^{(\ell )}[\kappa ]=\frac{\kappa }{2\pi }r_{p}[\kappa ,\kappa ]^{\ell
}-\int_0 ^{\kappa }\frac{{\rm d}\omega }{2\pi }\omega \partial _{\omega
}r_{p}[\omega ,\kappa ]^{\ell }
\]
we write 
\[
R_{p}^{(\ell )}[\kappa ]\approx \frac{\kappa }{2\pi }r_0 ^{\ell }\qquad
r_0 =\left\{ r_{p}[\kappa ,\kappa ]\right\} _{\kappa \rightarrow 0}
\]
The value $r_0 $ is evaluated at zero frequency and at normal incidence.
This value is the same for the two polarizations and one obtains finally the
total Casimir force as 
\begin{equation}
F=\sum_{\ell >0}r_0 ^{\ell }C(2\ell q)\qquad C(\tau )=\frac{6\hbar }{\pi
^2 \tau ^{4}}  \label{eq36}
\end{equation}
that is 
\begin{equation}
F=\frac{3\hbar }{8\pi ^2 q^{4}}\zeta _{r_0 }(4)  \label{eq37}
\end{equation}
where $\zeta _{x}(p)$ is defined by (\ref{eq21}). At the limit of perfectly
reflecting mirrors, one recovers the expression (\ref{eq2}) for the Casimir
force $\left( \zeta _1 (4)=\frac{\pi ^{4}}{90}\right) $.

As in the 2D case, the expressions of the force can be written in terms of
the reflectivities evaluated for imaginary frequencies (see the appendix).

A phase shift representation can be obtained for any distance by integrating
the force. One writes the energy ${\cal U}$ as a sum over the ordinary and
evanescent waves 
\[
{\cal U}=\int_0 ^{\infty }\frac{{\rm d}\omega }{2\pi }\int_{i\infty }^1 
\frac{{\rm d}\gamma }{2\pi }\frac{\hbar \gamma \omega ^2 }2 \left( -\Delta
[\omega ,\omega \gamma ]\right) 
\]
where $\Delta [\omega ,\kappa ]$ is obtained by summing all the phase shifts
at frequency $\omega $ and wavevector $\kappa $ 
\begin{eqnarray*}
\Delta [\omega ,\kappa ] &=&\Delta _{p_1 }[\omega ,\kappa ]+\Delta
_{p_2 }[\omega ,\kappa ] \\
\Delta _{p}[\omega ,\kappa ] &=&i{\rm Log}\frac{1-r_{p}[\omega ,\kappa
]e^{2i\kappa q}}{1-r_{p}[\omega ,\kappa ]^{*}e^{-2i\kappa q}}
\end{eqnarray*}
Moving the integration path leads to expressions in terms of the unphysical
waves 
\[
{\cal U}=\int_0 ^{\infty }\frac{{\rm d}\omega }{2\pi }\int_1 ^{\infty }
\frac{{\rm d}\gamma }{2\pi }\frac{\hbar \gamma \omega ^2 }2 \Delta [\omega
,\omega \gamma ]
\]
These expressions can be written in terms of $\partial _\omega \Delta
[\omega ,\omega \gamma ]$ by an integration by parts 
\begin{eqnarray*}
{\cal U} &=&\int_0 ^{\infty }\frac{{\rm d}\omega }{2\pi }\int_{i\infty }^1 
\frac{{\rm d}\gamma }{2\pi }\frac{\hbar \gamma \omega ^{3}}{6}\partial
_\omega \Delta [\omega ,\omega \gamma ] \\
&=&\int_0 ^{\infty }\frac{{\rm d}\omega }{2\pi }\int_1 ^{\infty }\frac{
{\rm d}\gamma }{2\pi }\frac{\hbar \gamma \omega ^{3}}{6}\left( -\partial
_\omega \Delta [\omega ,\omega \gamma ]\right) 
\end{eqnarray*}
Writing $\partial _\omega \Delta [\omega ,\omega \gamma ]$ as in equation 
(\ref{eq27}), one obtains the Casimir energy as the sum of the integrated
field energy and of corrections associated with the reflection delays upon
the mirrors. The integrated field energy is here 
\[
{\cal U}=-\frac{qF}{3}
\]
In the limit of a perfect reflection, this relation is compatible with the
definition (\ref{eq23}) of ${\cal U}$ \cite{Casimir7} since the force scales
as $q^{-4}$.

The force for non zero temperatures is given by equations (\ref{eq33}-\ref
{eq34}) with the function $c$ modified according to equation (\ref{eq28}).
It is also given by equation (\ref{eq35}) with 
\[
R_{p}^{(\ell )}[\kappa ]=\int_0 ^{\kappa }\frac{{\rm d}\omega }{2\pi }
r_{p}[\omega ,\kappa ]^{\ell }n_{T}[\omega ]
\]
At the large distance approximation, one obtains 
\[
F=\sum_{\ell >0}r_0 ^{\ell }C(2\ell q)\qquad C(\tau )=\frac{\hbar }{\pi ^2 }
\partial _{\tau }^2 \frac{\alpha }{\tanh (\alpha \tau )}\qquad \alpha =
\frac{\pi T}{\hbar }
\]
The low temperature limit gives the known result. For high temperatures, one
obtains by omitting exponentially small terms 
\begin{eqnarray}
&&C(\tau )\approx \frac{2T}{\pi \tau ^{3}}\quad {\rm for\ }T\tau \gg \hbar  
\nonumber \\
&&F=\frac{T}{4\pi q^{3}}\zeta _{r_0 }(3)
\end{eqnarray}
where $\zeta _{x}(p)$ is defined by (\ref{eq21}). High temperatures
correspond to a classical limit so that the expression does not contain 
$\hbar $. For a perfect mirror, one recovers the known result \cite{Casimir4}.

In the general case, one also obtains a phase shift representation of the
free energy 
\begin{eqnarray*}
{\cal F} &=&\int_0 ^{\infty }\frac{{\rm d}\omega }{2\pi }\int_{i\infty }^1 
\frac{{\rm d}\gamma }{2\pi }\frac{\hbar \gamma \omega ^2 }2 n_{T}[\omega
]\left( -\Delta [\omega ,\omega \gamma ]\right)  \\
&=&\int_0 ^{\infty }\frac{{\rm d}\omega }{2\pi }\int_1 ^{\infty }\frac{
{\rm d}\gamma }{2\pi }\frac{\hbar \gamma \omega ^2 }2 n_{T}[\omega ]\Delta
[\omega ,\omega \gamma ]
\end{eqnarray*}

\medskip
\noindent {\bf Acknowlegdements}

We thank C. Cohen-Tannoudji, C. Fabre, E. Giacobino, S. Haroche, A. Heidmann and
H.M. Nussenzveig for stimulating comments.

\appendix 

\section{Expressions of the forces as integrals over imaginary frequencies}

Following the computation by Lifshitz of the Casimir-Polder forces \cite
{Casimir8}, one can write the forces as integrals over imaginary frequencies.

In the 2D case, one first shows that $\left| r[\omega ]\right| $ remains
smaller than $1$ for any $\omega $ in the upper half plane $\Im \omega >0
$. The modulus${}\left| r[\omega ]\right| $ is less than $1$ on the real
axis (from the unitarity and reality conditions); it goes to $0$ for large
values of $\omega $ (from the transparency condition); as $r[\omega ]$ is
analytic for $\Im \omega >0$, the conclusion is a consequence of the
Phragm\'{e}n-Lindel\"{o}f theorem \cite{Casimir13}.

Therefore, the integrand appearing in (\ref{eq17}) has no poles in the upper
half plane. Using the transparency condition, one can then move the
integration path $\left[ 0,\infty \right[ $ to the upper part of the
imaginary axis $\left[ 0,i\infty \right[ $ 
\[
F=\int_0 ^{\infty }{\rm d}\omega \frac{\hbar \omega }{2\pi }\frac{r[i\omega
]}{e^{2\omega q}-r[i\omega ]}+c.c.
\]
Furthermore, $r[i\omega ]$ is real since it is the Laplace transform of the
causal and real function $r$ 
\[
r[i\omega ]=\int_0 ^{\infty }{\rm d}t\ r(t)e^{-\omega t}
\]
It follows that the Casimir force is 
\[
F=\int_0 ^{\infty }{\rm d}\omega \frac{\hbar \omega }{\pi }\frac{r[i\omega ]}
{e^{2\omega q}-r[i\omega ]}
\]
This equation can conveniently be used for evaluating the Casimir force for
any reflectivity functions 
\begin{equation}
F=\frac{\hbar }{4\pi q^2 }\int_0 ^{\infty }{\rm d}u\frac{ur\left[ \frac{iu}
{2q}\right] }{e^{u}-r\left[ \frac{iu}{2q}\right] }  \label{eq39}
\end{equation}
At the large separation limit, $r\left[ \frac{iu}{2q}\right] $ may be
replaced by $r_0 $ and one obtains an integral expression equal to the
series of equation (\ref{eq20}) 
\[
F=\frac{\hbar }{4\pi q^2 }\int_0 ^{\infty }{\rm d}u\frac{ur_0 }
{e^{u}-r_0 }
\]

The integral (\ref{eq39}) can be written as a formal series 
\begin{eqnarray*}
F &=&\sum_{\ell >0}F^{(\ell )} \\
F^{(\ell )} &=&\frac{\hbar }{\pi }\left\{ \left( 2\ell q-\partial _{\omega
}\right) ^{-1}\omega r[i\omega ]^{\ell }\right\} _{\omega =0}=\frac{\hbar }
{\pi }\left\{ \left( 2\ell q-\partial _\omega \right) ^{-2}r[i\omega
]^{\ell }\right\} _{\omega =0}
\end{eqnarray*}
This equation and the expression (\ref{eq22}) are formally equivalent;
however, they have a different status: $r[i\omega ]$ is real and plays the
role of a regulator for the mode density while $r[\omega ]$, which appears
in (\ref{eq22}), cannot be real because of causality and high frequency
transparency.

One can also write the 4D Casimir force as an integral over imaginary
frequencies. We start from the integral (\ref{eq33}) where the integration
path for $\gamma $ includes the ordinary contributions $\left( 0<\gamma
<1\right) $ as well as the evanescent waves ($\gamma $ imaginary) and we
treat separately these two contributions. For the ordinary contributions, it
follows from causality that: $r_{p}[\omega ,\omega \gamma ]$ is analytic in 
$\omega $ for $[ \Re \omega >0,\Im \omega >0]$ and $\gamma $ a fixed
real number; $\left| r_{p}[\omega ,\omega \gamma ]\right| \leq 1$ for $[
\Re \omega >0,\Im \omega >0]$ and $\gamma $ real. Using the high
frequency transparency, we move the integration path for $\omega $ to the
upper imaginary axis (with $\gamma $ kept constant). For the evanescent
waves, we use the analogous properties \cite{Casimir22}: $r_{p}[\omega
,\kappa ]$ is analytic in $\omega $ for $[\Re \omega >0,\Im \omega
>0]$ and $\kappa $ a fixed imaginary number; $\left| r_{p}[\omega ,\kappa
]\right| \leq 1$ for $[\Re \omega >0,\Im \omega >0]$ and $\kappa $
imaginary. Using the high frequency transparency, we move the integration
path for $\omega $ to the upper imaginary axis (with $\kappa $ kept
constant). Finally, we use the property: $r_{p}[\omega ,\kappa ]$ real for 
$\omega $ and $\kappa $ imaginary.

At the end of the process, one obtains 
\[
F_{p}=\frac{\hbar }{2\pi ^2 }\int_0 ^{\infty }{\rm d}\kappa \ \kappa
^2 \int_0 ^{\kappa }{\rm d}\omega \frac{r_{p}[i\omega ,i\kappa ]}
{e^{2\kappa q}-r_{p}[i\omega ,i\kappa ]}
\]
At the large separation limit, $r_{p}[i\omega ,i\kappa ]$ is replaced by a
constant $r_0 $ and one obtains the Lifshitz' integral expression \cite
{Casimir8} equal to the series of equation (\ref{eq37}) 
\[
F=\frac{\hbar }{32\pi ^2 q^{4}}\int_0 ^{\infty }{\rm d}u\frac{u^{3}r_0 }
{e^{u}-r_0 }
\]


\begin{references}
\bibitem{Casimir20}  One can alternatively consider that the elementary
impulsion transfer is multiplied by $\gamma $ as well as the cross section
of a given plate area \cite{Casimir5}.
\bibitem{Casimir21}  This corresponds to a property often encountered in
scattering theory since $\kappa $ is the momentum transfer in the reflection 
\cite{Casimir13}.
\bibitem{Casimir22}  These properties are satisfied in the particular case
of interfaces between dielectric media \cite{Casimir8}.
\bibitem{Casimir23}  More precisely, we assume that $\left| r_{p}[\omega
,\kappa ]\right| $ is less than $1$ for the ordinary waves (from the
unitarity condition), for the evanescent waves and for the unphysical waves $%
\omega <\kappa $; it follows from the Phragm\'{e}n-Lindel\"{o}f theorem \cite
{Casimir13} that $\left| r_{p}[\omega ,\kappa ]\right| $ is less than $1$ in
the quadrant.
\end{references}
\end{document}